\documentclass[fleqn,12pt,twoside]{article}
\usepackage{espcrc1}\usepackage{wrapfig}
\usepackage{graphicx}\usepackage{epsfig}
\newcommand{\be}{\begin{eqnarray}}

\newcommand{\ee}{\end{eqnarray}}

\newcommand{\bb}{b_\perp}
\newcommand{\rr}{r_\perp}

\makeindex

\newcommand{\beq}{\begin{eqnarray}}
\newcommand{\eeq}{\end{eqnarray}}
\def\labe{\label}
\def\simge{\mathrel{%
   \rlap{\raise 0.511ex \hbox{$>$}}{\lower 0.511ex \hbox{$\sim$}}}}
\def\simle{\mathrel{
   \rlap{\raise 0.511ex \hbox{$<$}}{\lower 0.511ex \hbox{$\sim$}}}}
\def\bigs{\mathrel{
   \rlap{\raise 0.531ex \hbox{$>$}}{\lower 0.531ex \hbox{$<$}}}}

\def\del{\partial}                              

\newcommand{\kk}{k_\perp}

\title{{\bf The Colour Glass 
Condensate}}

\author{Edmond Iancu\address{Service de Physique 
Th\'eorique, CEA Saclay,
           91191 Gif-sur-Yvette cedex, France}}
       
\begin{document}

\maketitle
\setcounter{footnote}{0}

\begin{abstract}
I review the physical and mathematical foundations for
the theoretical description of the hadron wavefunction at small $x$ as
a Colour Glass Condensate. In this context, I discuss the phenomenon of gluon 
saturation and some of its remarkable
consequences: a new ``geometric scaling'' for $F_2$,
which has been recently identified at HERA, and the unitarization of the hadronic
cross-sections at high energy. I show that by combining saturation and confinement
one obtains cross-sections which saturate the Froissart bound.
\end{abstract}

\section{INTRODUCTION} 

Understanding the high-energy behaviour of hadronic cross-sections is a most
intriguing and fascinating problem, which is intimately related to the physics
of high parton densities. Linear evolution equations in QCD, 
like DGLAP and BFKL, predict a rapid growth of the parton densities, 
which violates unitarity constraints. This is the ``small--$x$
problem''\footnote{As usual, $x$ denotes the fraction of the hadron longitudinal
momentum carried by a typical parton involved in the scattering: $x\sim Q^2/s$,
with $Q^2$ a typical transferred momentum. Thus, the high energy limit
($s \to\infty$ at fixed $Q^2$) is the same as the 
small--$x$ limit.} of perturbative QCD. But the applicability of
perturbation theory to high-energy problems is by no means obvious.
Indeed, even when $s \to\infty$ (with $s$ the total energy squared
in the center of mass frame), there is a priori no guarantee that all the 
transferred momenta are large. 
The ``infrared diffusion'' problem of the BFKL equation \cite{BFKL}
(see Sect. 2.1 below)
provides an explicit counterexample in this respect.

But this does not prove the inadequacy of perturbation theory in general. 
In fact, the strong rise of the parton densities carries in itself
the potential for its own solution \cite{GLR,MQ86,BM87,MV94}:
The interactions between the partons from different parton cascades,
 omitted in the linear evolution equations, will become more
and more important with increasing energy, and may eventually tame the growth
of the parton densities. This {\it parton saturation} phenomenon
is expected to introduce a characteristic momentum scale,
the {\it saturation momentum} $Q_s(x,A)$, which is a measure of the
density of the saturated {\it gluons}, and grows rapidly with $1/x$ and 
$A$ (the atomic number). 
Thus, for sufficiently high energies and/or very large nuclei,
gluon saturation provides
an intrinsic hard momentum scale which limits the ``infrared diffusion'' 
and justifies the use of weak coupling methods: $\alpha_s(Q_s) \ll 1$.

Note, however, that although the coupling is weak, the saturation regime
remains non-perturbative in that the non-linear effects associated with
the high parton densities cannot be expanded out in some perturbative scheme,
but must be included exactly. This is a priori a formidable task, but progress can be
made by recognizing that this is a {\it semi-classical} regime,
for which a classical {\it effective theory} can be written \cite{MV94}.
This is a theory for the hadron wavefunction in the
infinite momentum frame, and it looks remarkably
simple: It is just classical Yang-Mills theory in the presence of a 
random distribution of colour charges which propagate nearly at the speed of light,
and whose internal dynamics is ``frozen'' by Lorentz time dilation.
The classical gauge fields represent the ``small--$x$
gluons'', i.e., the gluons with small longitudinal momenta, which are
the relevant degrees of freedom for high-energy scattering.
The fast moving charges represent the partons with relatively high longitudinal 
momenta, like the valence quarks, which are not directly involved in the
scattering, but act as sources for the small--$x$ gluons.
At saturation, the colour fields are strong, and the classical theory must be
solved exactly.

Originally, this theory has been formulated as a model to study
saturation effects in a large nucleus ($A \gg 1$) at not so high
energies \cite{MV94,K96,JKMW97,KM98,LM00}. In that case, the gluon
density is high because of the presence of many colour sources (the $A\times N_c$ 
valence quarks) already at ``tree-level''.
Subsequently, it has been shown that this effective theory is consistent
with the QCD evolution towards higher energies
\cite{JKLW97,PI}: The evolution modifies the classical colour source
by incorporating the quantum gluons with longitudinal momenta above the
small--$x$ scale of interest. This involves a renormalization
group equation which can be seen as a functional non-linear generalization
of the BFKL equation (see Sect. 3 below).

We have dubbed the gluonic matter described by this effective theory
a {\it Colour Glass Condensate} (CGC) \cite{PI,Cargese}:
\vspace*{-.1cm}
\begin{itemize}
\item {``{\it Colour }'' since gluons are ``coloured''
under SU(3).}\vspace*{-.2cm}
\item {``{\it Glass }'' because of the analogy with the mathematical
description of systems with frozen disorder, called ``glasses''.
Here, the ``frozen disorder'' refers to the random distribution of
time-independent colour charges, which is averaged over in the calculation
of physical observables.}
\vspace*{-.2cm}
\item {``{\it  Condensate }'' since, at saturation, the gluon modes
have large occupation numbers, of  order $1/\alpha_s$ (corresponding
to strong classical fields $A\sim 1/g$), which is the maximal
value allowed by the repulsive gluon interactions.
This is a Bose condensate.}
\end{itemize}
\vspace*{-.2cm}
It is my purpose in this talk to review the physical and mathematical foundations
of the CGC picture, and describe some of its
consequences for gluon saturation \cite{AM2,SAT}, geometric scaling 
in deep inelastic scattering 
\cite{geometric,GS2,SCALING}, and the unitarization of hadronic cross-sections
\cite{FB}. I will also discuss ``limiting fragmentation'' as an incentive
towards the renormalization group description of multiparticle production
in heavy ion collisions.
Other consequences and recent applications to the phenomenology of 
deep inelastic scattering and heavy ion collisions will be briefly mentioned
towards the end. More detailed discussions can be found in specialized
contributions to these proceedings \cite{QM02}.

\section{FROM BFKL TO GLUON SATURATION}
\setcounter{equation}{0}

\begin{figure}[htb]\vspace*{-.2cm}
\centering
\resizebox{.8\textwidth}{!}{%
\includegraphics*{{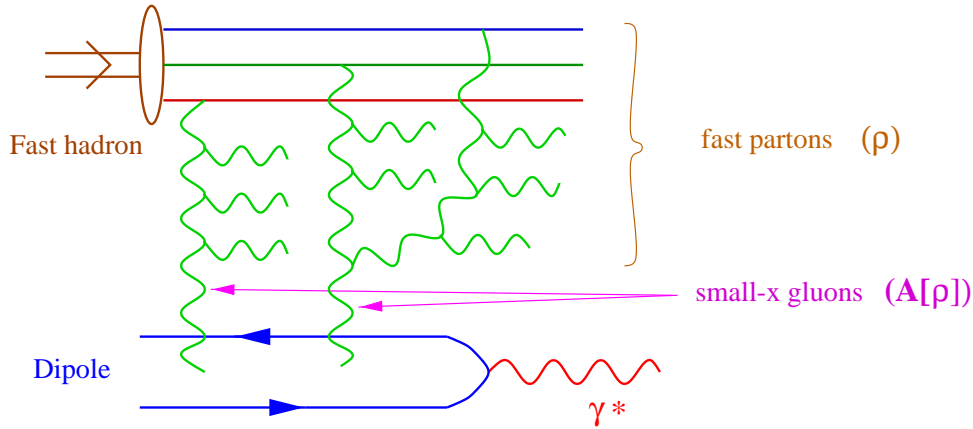}}}
\vspace*{-.3cm}\caption{Deep inelastic scattering in the dipole frame}
\vspace*{-.6cm}\label{DIS2}
\end{figure}

Consider deep inelastic scattering, for simplicity. We are interested in
the high-energy, or small--$x$, limit, with $x\simeq Q^2/s\ll 1$.
It is convenient to regard this process in a special frame (the dipole frame) in
which most of the total energy is carried by the hadron,
while the virtual photon $\gamma^*$ 
has just enough energy to dissociate before scattering into a quark--antiquark 
pair (a {\it colour dipole}), which then scatters off the gluon fields in the
hadron. (See Fig. \ref{DIS2}.)  In this frame, the DIS cross-section can be 
related to the dipole-hadron scattering amplitude ${\cal N}_\tau(r_\perp)$
\cite{AM0}, which to lowest order (two gluon exchange)
is in turn proportional
to the {\it gluon distribution} in the  hadron wavefunction:
\be\label{nss}
{\cal N}_\tau(r_\perp)\,\simeq\,r_\perp^2\,
\frac{\pi^2\alpha_s C_F}{N_c^2-1}\,\frac{x G(x,1/r_\perp^2)}{\pi R^2}\,
.\ee
Here, $r_\perp=x_\perp-y_\perp$ is the transverse size of the dipole
(with the quark at $x_\perp$ and the antiquark at $y_\perp$),
$\tau\equiv \ln(1/x)$ is the rapidity gap between $\gamma^*$ and the hadron
(which is almost the same as the hadron rapidity in the considered frame),
and $x G(x,Q^2)/\pi R^2$ is the number of gluons with 
longitudinal momentum fraction $x$ and transverse
size $r_\perp =1/Q$ per unit rapidity per unit transverse area.
Throughout, I assume  that the dipole is sufficiently small
to be perturbative: $Q^2\equiv 1/r_\perp^2 \gg \Lambda_{QCD}^2$.

In writing eq.~(\ref{nss}), I assumed the hadron to be  
homogeneous in the transverse
plane; that is, the gluon density was taken to be the same at all the
impact parameters within the hadron disk of radius $R$. 
(A more realistic impact parameter dependence will be considered in Sect. 7.)
Note that the scattering amplitude (\ref{nss}) is not sensitive to the details
of the longitudinal distribution of gluons inside the hadron, but only to their
projected distribution in the transverse plane. Physically, it is so since the
low-energy virtual photon has a large longitudinal wavelength $\propto 1/x$,
so it scatters coherently off all the partons in a longitudinal tube of transverse
size $r_\perp$.

The $F_2$ data at HERA show a rapid growth of the gluon distribution
with $1/x$, in qualitative agreement with the predictions of perturbative QCD
to which I now turn.

\subsection{The small--$x$ problem of the BFKL approximation}

\begin{figure}[htb]
\centering
\resizebox{.85\textwidth}{!}{%
\includegraphics*{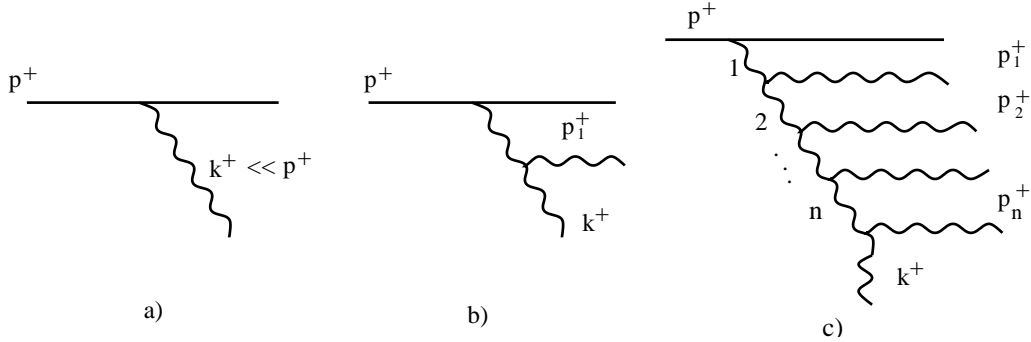}}\vspace*{-0.5cm}
   \caption{a) Small--$x$ gluon emission by a fast parton;
b) the lowest-order radiative correction; c) a gluon cascade.}
\label{gluoncascade}\vspace*{-0.3cm}
\end{figure}

\vspace*{0.2cm}
Within perturbative QCD, the enhancement of the gluon distribution at small $x$ 
proceeds via the gluon cascades depicted in Fig.~\ref{gluoncascade}.
Fig.~\ref{gluoncascade}.a shows
the direct emission of a soft gluon with longitudinal momentum\footnote{I use 
light-cone vector notations: $k^\pm=(k^0\pm k^3)/\sqrt{2}$, 
$k_\perp=(k^1,k^2)$. The total 4-momentum of the  hadron reads
$P^\mu=\delta^{\mu +}P^+$ with large $P^+$.}  $k^+ = xP^+
\ll P^+$ by a fast moving parton (say, a valence quark) with $p^+ = x_0P^+$
and $1> x_0\gg x$. Fig.~\ref{gluoncascade}.b displays the lowest-order
radiative correction
which is of the order (with $N_c$ the number of colours,
and $\bar\alpha_s\equiv\alpha_s N_c/\pi$)\vspace*{-0.9cm}
\beq\label{one-gluon}
\frac{\alpha_s N_c}{\pi}
\int_{k^+}^{p^+}\,\frac{d p_1^+}{p_1^+}
\,=\,\frac{\alpha_s N_c}{\pi}
\,\ln\,\frac{p^+}{k^+}\,= \,\bar\alpha_s\ln\frac{x_0}{x}\,
\eeq
relative to the tree-level process in Fig.~\ref{gluoncascade}.a.
This correction is enhanced by the large { rapidity} interval
$\Delta \tau= \ln(x_0/x)$ available for the emission of the additional
gluon. A similar enhancement holds for the gluon cascade  
in Fig.~\ref{gluoncascade}.c, in which the succesive gluons are strongly ordered
in longitudinal momenta: $p^+\gg p_1^+\gg p_2^+ \gg\,\cdots\,\gg 
p_n^+\gg k^+$. This gives a contribution of relative order
$\frac{1}{n !}
\left(\bar\alpha_s\ln\frac{x_0}{x}\right)^n\,.$
Clearly, when $x$ is so small that $\ln(x_0/x)\sim 1/\bar\alpha_s$,
all such quantum ``corrections'' become of order one, and must be
resummed for consistency. This gives the 
number of gluons with longitudinal momentum fraction $x$
(or rapidity $\tau=\ln(1/x)$) within this ``leading-log'' approximation:
\beq\label{expgr}
{dN\over d\tau}\,\equiv \,x G(x,Q^2)\,\sim\,
{\rm e}^{\omega \bar\alpha_s\tau}\,=\,
{x^{-\omega \bar\alpha_s}}\,,\eeq
with $\omega$ a pure number. We have tacitly assumed that all the gluons
in the cascade have transverse momenta of the same order, namely of
order $Q$. A more refined treatment, based on the BFKL equation \cite{BFKL}, 
allows one to compute $\omega$ and specifies the $Q^2$--dependence of the
distribution. One obtains:
\be\label{GBFKL}
\frac{x G(x,Q^2)}{\pi R^2}\bigg|_{BFKL}\,\simeq\,
\sqrt{{\Lambda^2}{Q^2}}\,
\, \exp\Bigg\{\omega\bar\alpha_s\tau
-\frac{1}{2\beta\bar\alpha_s\tau}\left(\ln \frac{Q^2}{\Lambda^2}\right)^2
\Bigg\},\ee
where $\omega\!=\!4\ln 2\approx 2.77$, $\beta\!=\!28\zeta(3)\approx 33.67$,
and $\Lambda$ is a reference scale, of order $\Lambda_{QCD}$. 
Eq.~(\ref{GBFKL}) exhibits two essential features of the gluon distribution
in the BFKL approximation: ({\it a}) its exponential increases 
with $\tau$, and
({\it b}) its diffusive behaviour in $\ln ({Q^2}/{\Lambda^2})$, with diffusion
``time'' equal to $\tau$. These features are responsible for the main
difficulties of this approximation in the high energy limit :
\begin{description}
\item{ ({\it a}) {\it Violation of the unitarity bound}} :
At high energy, hadronic cross-sections are proportional to the gluon distribution
(see, e.g., eq.~(\ref{nss})). Then, eq.~(\ref{GBFKL}) predicts
$\sigma_{\rm tot}(s)\sim s^{\omega \bar\alpha_s}$, which violates the
Froissart bound $\sigma_{\rm tot}(s) \le \sigma_0\ln^2 s$ \cite{Froissart}.
\item{ ({\it b}) {\it Infrared diffusion}} : With increasing $s$, the typical
transverse momenta carried by the gluons within the BFKL ladder
diffuse into the non-perturbative region at $Q^2 \simle \Lambda^2$.
This contradicts the use of perturbation theory.
\end{description}

\subsection{The saturation momentum}

\begin{figure}[htb]
\centering
\resizebox{.5\textwidth}{!}{%
\includegraphics*{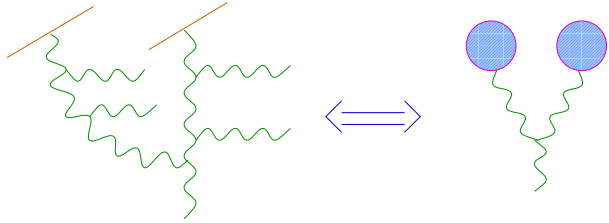}}\vspace*{-.3cm}
\caption{Gluon recombination in the hadron wavefunction.}
\label{GLRfig}\vspace*{-.5cm}
\end{figure}

\vspace*{0.2cm}
When resumming the ladder diagrams as above, we have assumed that 
the emitted gluons propagate freely, that is, we have neglected
their interactions with gluons at the same rapidity from other
cascades (cf. Fig. \ref{GLRfig}).
Such interactions are not enhanced by a large logarithm, but may
nevertheless become important when the density of the
available gluons is large enough.
Specifically, the interaction probability for gluons from different
parton cascades can be estimated as:
\be\label{wglr}
\sigma(Q^2)\times n(x,Q^2)\,\sim\,
\frac{\alpha_s N_c}{Q^2}\,\,\times\,\frac{1}{N_c^2-1}\frac{x G(x,Q^2)}{\pi R^2}
\ee
where $\sigma(Q^2)$ is a typical cross-section for 
gluons with transverse size $1/Q$, and $n(x,Q^2)$ is the density
of the gluons of a given colour, and with given $x$ and $1/Q$, in the transverse plane.
This probability becomes of order one for $Q^2$ lower than
a critical value :
\be\label{Qs}
Q^2_s(\tau)\,\simeq\,\frac{\alpha_s N_c}{N^2_c-1}\,\frac{x G(x,Q^2_s)}{\pi R^2}\,,\ee
that we shall refer to as the {\it saturation momentum}. 
More precisely, this is the scale at which non-linear effects start to be 
important in the hadron wavefunction:

For $Q^2\gg Q^2_s(\tau)$, the non-linear effects
are negligible, and linear evolution equations (like BFKL or DGLAP) apply.
In particular, one can estimate the saturation scale by
inserting the BFKL approximation (\ref{GBFKL}) into
eq.~(\ref{Qs}). This gives \cite{AM2,SCALING} :
\be\label{Qs0}
Q_s^2(\tau)\,=\,\Lambda^2 
{\rm e}^{c\bar\alpha_s \tau}\,,\qquad c\,=\,
\big[-{\beta}+\sqrt{\beta(\beta+8\omega)}\,\big]/2\,=\,4.84...\ee

For $Q^2\simle Q^2_s(\tau)$, the non-linear effects 
are essential, and are expected to tame the growth of the gluon distribution with
$\tau$. It has been predicted almost twenty years ago \cite{GLR} that a ``saturation''  
regime should be reached in which recombination equilibrates radiation 
(see also Refs. \cite{MQ86,BM87,MV94}).
To verify this conjecture, a formalism is needed to perform
calculations in the non-linear regime. This will be described in Sect. 3.

For a nucleus, $x G_A(x,Q^2_s)\propto A$ and 
 $\pi R^2_A\propto A^{2/3}$, so eq.~(\ref{Qs}) predicts 
$Q^2_s\propto A^{1/3}$.
To summarize 
(with $\delta\approx 1/3$ and $\lambda\approx c\bar\alpha_s$ in a first
approximation) :
\be\label{QsxA}
Q^2_s(x,A)\,\,\sim\,\,A^{\delta}\, x^{-\lambda}\,,\ee
which shows that an efficient way to create a high-density
environment is to combine large nucleai with moderately small values of $x$,
as currently done at RHIC.
Schematically:
\begin{description}
\vspace*{-0.1cm}
\item --- $ep$ at HERA $\sim$ heavy ions at RHIC
\vspace*{-0.1cm}
\item --- $eA$ at eRHIC $\sim$ heavy ions at LHC
\vspace*{-0.1cm}
\end{description}
Eq.~(\ref{QsxA}) also shows that the saturation momentum 
becomes a {\it hard} scale ($Q^2_s(x,A)\gg \Lambda^2_{QCD}$) for
sufficiently large nuclei and/or large enough energies.
Once this condition is satisfied, we can reliably use weak coupling techniques
to study the evolution towards even higher energies. This strongly
suggests that the high energy limit
of QCD should be within the reach of perturbation theory.

\subsection{Saturation vs. Unitarization}

\vspace*{0.2cm}
Since unitarity violations by the BFKL approximation are related to the
unlimited growth of the gluon distribution, we expect that, converserly,
saturation effects should restore unitarity for hadronic cross-sections at
high energies.
Consider, e.g., the dipole-hadron scattering amplitude, for which the
unitarity bound reads ${\cal N}_\tau(r_\perp)\le 1$. (The upper limit
${\cal N}_\tau(r_\perp)= 1$ corresponds to ``blacknesss'': the dipole
is completely absorbed.) Of course, BFKL violates this bound for high enough
energies. But a brief inspection of  eqs.~(\ref{nss}) and (\ref{Qs})
reveals that the unitarity violations show up precisely at the scale at which
BFKL is expected to break down because of non-linear effects:
\be\label{Nsat}
{\cal N}_\tau(r_\perp)\,\simeq\, 1\qquad{\rm for}\quad r_\perp\,\sim\,
1/Q_s(\tau)\,.\ee
This suggests that unitarization effects and saturation are different 
aspects  of the same non-linear physics, and  cannot be dissociated from
each other. We shall see below that, indeed, the  non-linear effects
responsible for saturation cure the unitarity problem as well.

\section{THE EFFECTIVE THEORY FOR THE CGC} 
\setcounter{equation}{0}

According to eq.~(\ref{Qs}), the gluon distribution is large at saturation:
$xG(x,Q^2) \sim 1/\alpha_s$ for $Q^2\simle Q^2_s(\tau)$, 
which suggests the use of semi-classical methods.
One can write a classical effective theory whose general structure
is fixed by the kinematics of the infinite momentum frame \cite{MV94}:
The ``fast partons'' move nearly at the speed of light in the positive
$z$ (or positive $x^+$) direction, and generate a colour current 
$J^\mu_a=\delta^{\mu +}\!\rho_a$. By Lorentz contraction, the support
of the charge density $\rho_a$ is concentrated near $z=t$, or
$x^-=0$. By Lorentz time dilation, $\rho_a$ is independent of the
light-cone time $x^+$. 

The classical equation of motion (EOM) reads 
therefore:
\beq
(D_{\nu} F^{\nu \mu})_a(x)\, =\, \delta^{\mu +} \rho_a(x^-,x_\perp)\,.
\label{cleq}
\eeq
Physical observables are obtained by averaging the solution
to this equation over all the configurations of $\rho$, 
with a gauge-invariant weight function $W_\tau[\rho]$ which depends 
upon the dynamics of the fast modes. For instance, the
dipole-hadron scattering amplitude is computed as (in the eikonal
approximation \cite{BH,B}) :
\beq\label{Ntau}
{\cal N}_\tau(x_{\perp},y_{\perp})&=&1\,-\,\frac{1}{N_c}\,
\Big\langle {\rm tr}\Big(V^\dagger(x_{\perp}) V(y_{\perp})\Big)
\Big\rangle_\tau,\nonumber\\
\Big\langle {\rm tr}\Big(V^\dagger(x_{\perp}) V(y_{\perp})\Big)
\Big\rangle_\tau&=& \int { D}\rho\, \,W_\tau[\rho]\,\,{\rm tr}\Big
(V^\dagger_{x_{\perp}}[\rho] V_{y_{\perp}}[\rho]\Big),\eeq
where $V^\dagger(x_{\perp})$ and $V(y_{\perp})$ are Wilson lines
describing the interactions between the fast moving quark, or
antiquark, from the dipole and the colour field in the hadron:
\be\labe{v}
V^\dagger(x_{\perp})\,\equiv\,{\rm P} \exp
 \left \{ig \int dx^-\,A^+_a (x^-,x_{\perp})T^a
 \right \},\ee
and $A^+[\rho]$ is the solution to the EOM (\ref{cleq}) in the covariant
gauge: $- \nabla^2_\perp A^+ =\rho$.
The average over $\rho$ in 
eq.~(\ref{Ntau}) is similar to that performed for a
{\it spin glass} \cite{Cargese}.

As already mentioned, this ``glassy'' description is an {\it effective} theory, 
that is, it holds for a given (small) value of $x$, i.e., for a given longitudinal
momentum $k^+=xP^+\ll P^+$. The source $\rho_a$ is the  colour charge density
of the ``fast partons'' (the valence quarks and the gluons with momenta
$p^+ \gg k^+$ in the parton  cascades in Fig.~\ref{gluoncascade}). 
The classical solution $A[\rho]$ represents the
last gluon (with momentum $k^+$) in these cascades. 
The non-linear effects in the classical
EOM describe gluon recombination at the soft scale
$k^+$ (see Fig. \ref{GLRfig}). 
The corresponding effects at momenta $p^+ \gg k^+$ 
are included in the definition of $\rho$, that is, in the 
weight function $W_\tau[\rho]$.

All the dependence upon the
separation scale $k^+=xP^+$ is carried by the weight function
(via $\tau\equiv \ln(1/x)$) : When $x$ is further
decreased, say to $x_1 < x$, new quantum modes are effectively ``frozen''
(those with longitudinal momenta $x_1P^+ < p^+ < xP^+$),
and must be included in the effective source at the lower scale
$x_1P^+$. This can be done via a one-loop background field calculation, and
leads to a renormalization group equation for $W_\tau[\rho]$
which shows how the correlations of $\rho$ change with increasing $\tau$
\cite{JKLW97,PI}. Schematically:
\be\label{RGE}
{\del W_\tau[\rho] \over {\del \tau}}\,=\,
 {1\over 2} \int_{x_\perp,y_\perp}\,{\delta \over {\delta
\rho_\tau^a(x_\perp)}}\,\chi_{ab}(x_\perp, y_\perp)[\rho]\, 
{\delta \over \delta \rho_\tau^b(y_\perp)}\,W_\tau[\rho]\,,\ee
which describes diffusion in the functional space spanned by
$\rho_a(x^-,x_\perp)$ \cite{path}. The kernel 
$\chi[\rho]$, which plays the role of the
``diffusion coefficient'', is positive definite and non-linear in $\rho$ 
to all orders
(see  Refs. \cite{PI,Cargese} for an explicit expression and more details).

Eq.~(\ref{RGE}) resums both the large energy logarithms, i.e., the terms
$\sim (\alpha_s\ln 1/x)^n$, and the leading high density effects, i.e.,
the non-linear effects which become of order one at saturation.
Approximate solutions to this equation have been  obtained in Ref. \cite{SAT},
and will be described in Sect. 5 below. An exact, but formal, solution
in the form of a path-integral has been constructed in Ref. \cite{path},
where the stochastic nature of this equation has been fully elucidated,
and the equivalent Langevin equation (describing a random walk on a group
manifold) has been written down. Both the path-integral and the Langevin
formulation are well suited for numerical simulations on a lattice \cite{RW}.

Eq.~(\ref{RGE}) is a functional equation, but can be transformed
into ordinary evolution equations for the physical observables of interest.
For instance, by taking a derivative w.r.t. $\tau$ in eq.~(\ref{Ntau}),
and using eq.~(\ref{RGE}) for $\del W_\tau/\del \tau$, one can obtain
an equation for the  evolution of the scattering amplitude with $\tau$
\cite{PI}.
Note, however, that in general this is not a closed equation
(the 2-point function of the Wilson lines is coupled to a 4-point
function), 
but only the first step in an infinite hierarchy of coupled equations.
A closed equation for ${\cal N}_\tau(x_{\perp},y_{\perp})$ can be
nevertheless obtained in the large--$N_c$ limit. This is known
as the Balitsky-Kovchegov (BK) equation \cite{B,K}, and reads 
schematically\footnote{A similar non-linear equation
was originally suggested in Ref. \cite{GLR}
and proved in \cite{MQ86} in the double-log approximation. More recently,
Braun has reobtained this 
equation by resumming ``fan'' diagrams \cite{B00}.}:
\be\label{BK}
{\del  {\cal N}_\tau\over {\del \tau}}\,=\,\bar\alpha_s
K_{BFKL}\otimes {\cal N}_\tau
\,-\,\bar\alpha_s{\cal N}_\tau\otimes {\cal N}_\tau\,,\ee
that is, in addition to the linear BFKL terms, it contains also a
quadratic term which enforces unitarization (see Sect. 7).

So far, the picture of the quantum evolution has been developed
fully in the hadron infinite momentum frame: In the language of
Fig. \ref{DIS2}, the dipole rapidity has been kept fixed, while the
hadron has been accelerated to higher and higher energies, with the
result that the hadronic wavefunction (described here as a CGC)
evolves according to eq.~(\ref{RGE}).
Alternatively, via a change the frame, one could
use the increase in the total energy to accelerate the {\it dipole},
and study the evolution of its wavefunction with $\tau$.
In the language of eq.~(\ref{Ntau}), this would correspond to keeping
unchanged the weight function $W_{\tau_0}[\rho]$ ($\tau_0$ is the
 rapidity of the hadron which is now fixed), 
but modifying the scattering operator 
to allow for additional gluons in the (evolved) dipole wavefunction.
Since physics is boost invariant, both descriptions should 
lead to the same evolution equations for physical observables like
${\cal N}_\tau(x_{\perp},y_{\perp})$. And, indeed, the evolution
equations for Wilson line correlators derived from the CGC
\cite{PI} (in particular, eq.~(\ref{BK})) 
are identical to the equations originally obtained 
by Balitsky \cite{B} and Kovchegov \cite{K} from studies
of the dipole evolution. It has been first recognized by Weigert
\cite{W} that Balitsky's hierarchy of coupled equations
can be compactly summarized into a single functional equation, which
is essentially equivalent to eq.~(\ref{RGE}). More recently,
(See Ref. \cite{path} for a comparison between the two approaches.)

\section{LIMITING FRAGMENTATION}
\setcounter{equation}{0}

\begin{figure} 
\begin{center} 
\centerline{\epsfig{file=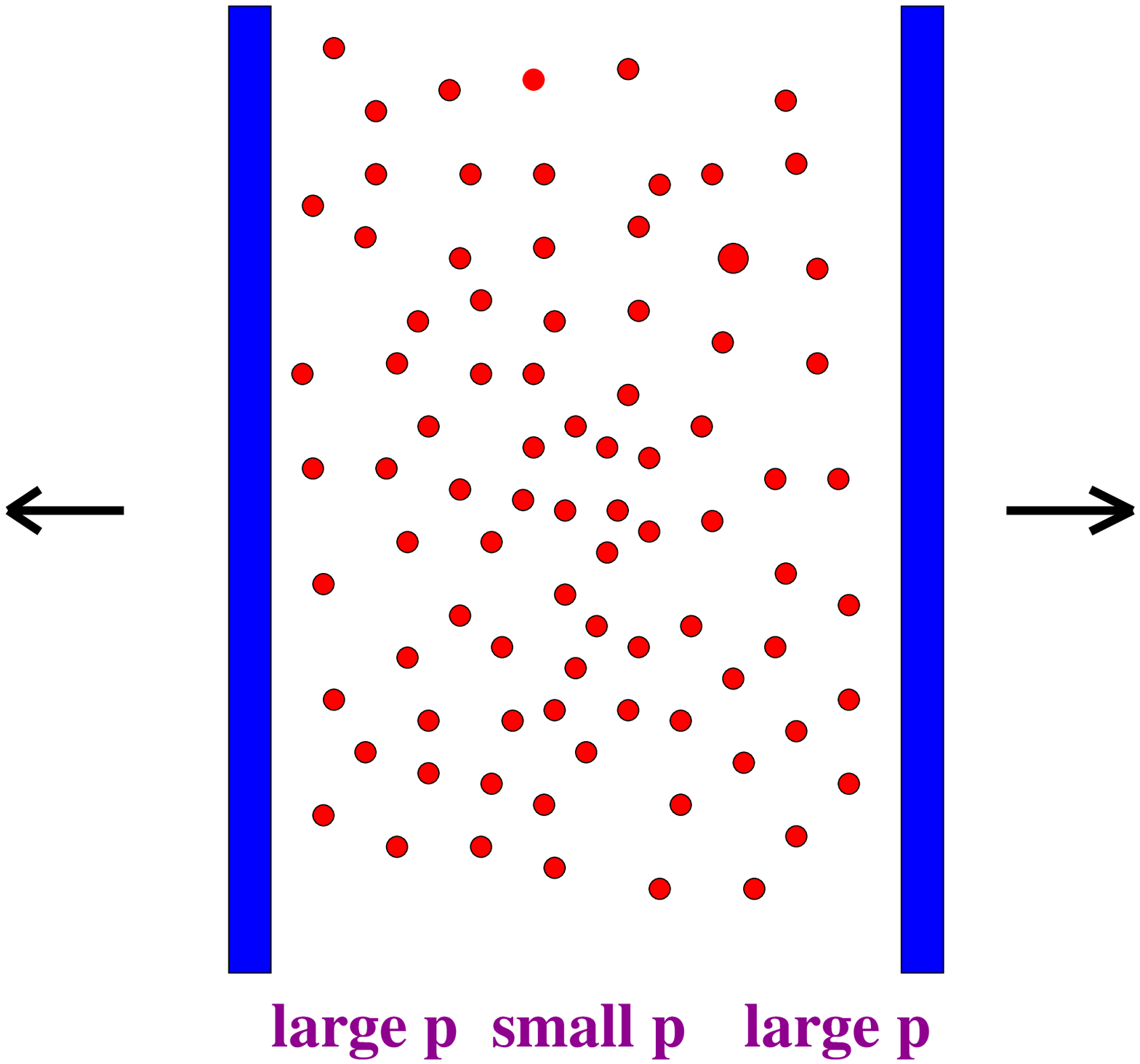,height=4.cm}
\epsfig{file=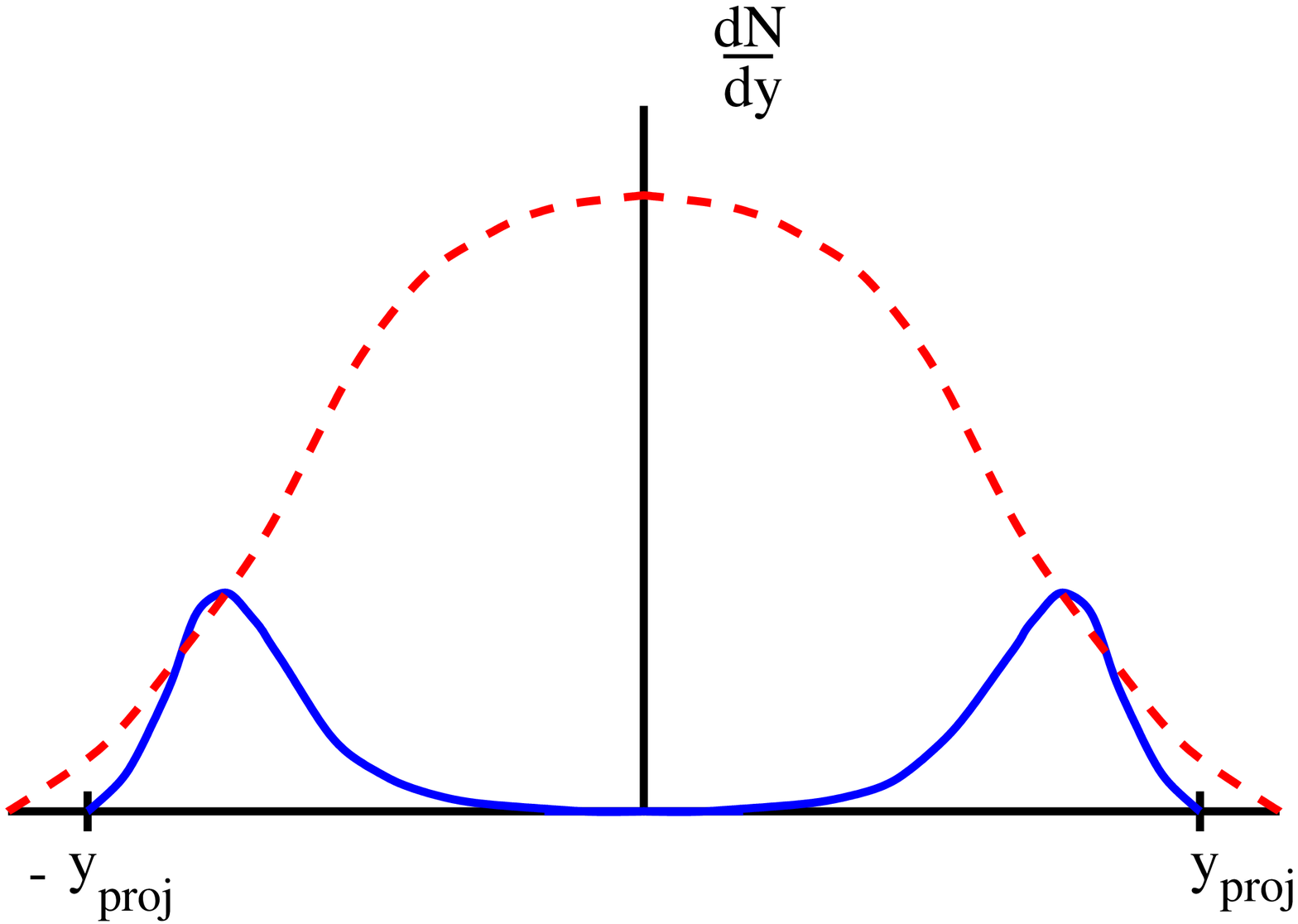,height=4.cm}
}
\caption{Particle production in heavy ion collisions.}
\end{center}\vspace*{-1.5cm}
\end{figure}
The  renormalization group description presented above is based on the
separation of scales in rapidity ($\ln 1/x$) for the degrees of freedom 
in the hadron wavefunction. This way to formulate the problem is
certainly useful for DIS, where one can tune the kinematics to observe
partons with a given value of $x$. Here, I would like to give you an
example which shows that the same strategy can be also
useful for hadron--hadron collisions, where partons with
all values of $x$ are a priori involved in the scattering.
 
Consider multiparticle production in, say, heavy ion collisions at RHIC.
A typical rapidity
distribution of the particles emerging from such a collision is
shown in Fig. 4. The leading particles are shown by the solid line 
and are clustered around the projectile and target rapidities.  
The dashed line is the distribution of the produced particles. 
Remarkably, this latter appears to reflect the rapidity distribution 
of the partons in the colliding hadron wavefunctions. 
In particular, it preserves the separation of scales in rapidity.
\begin{figure} 
\begin{center}\includegraphics[width=0.6\textwidth]
{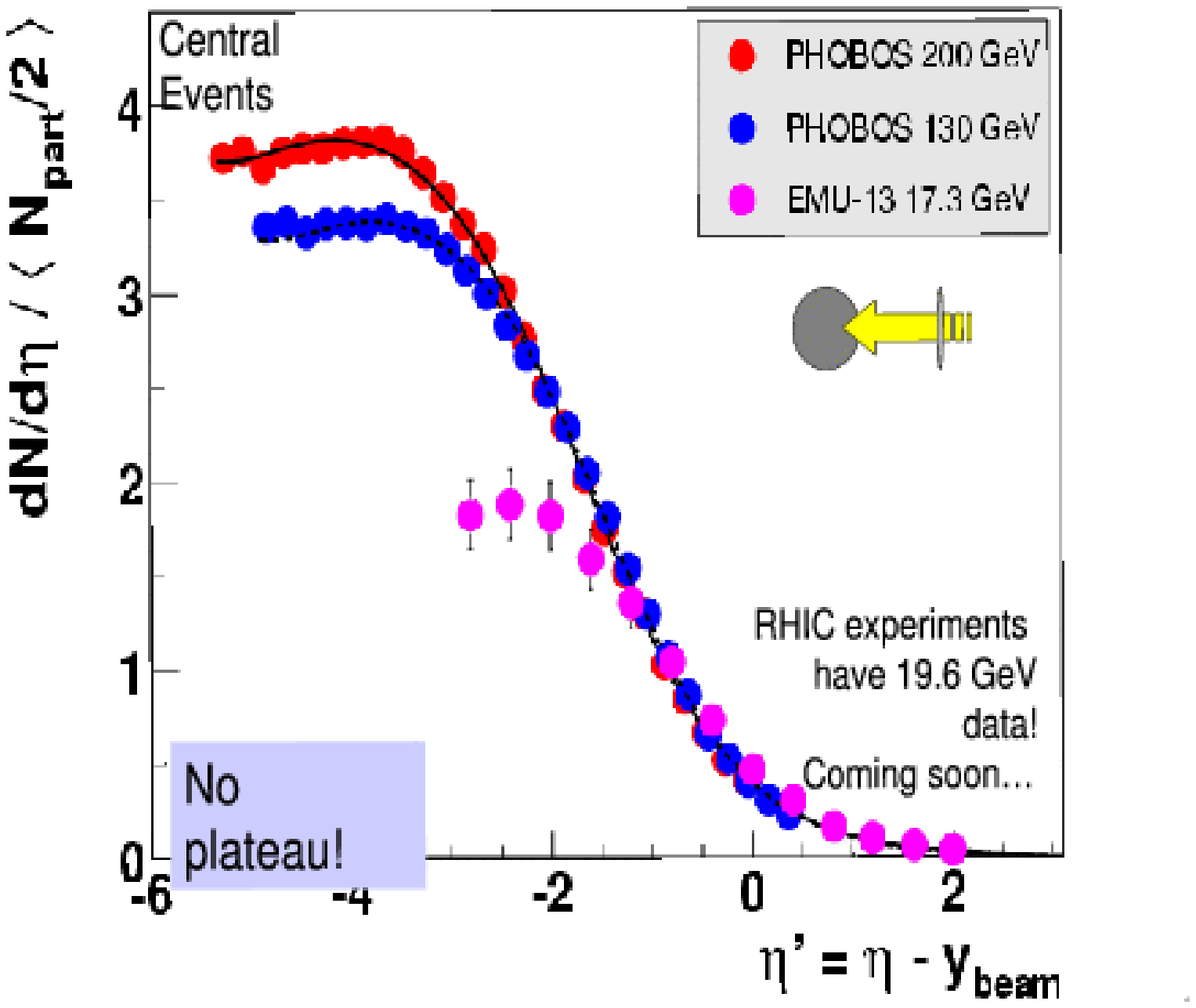} \vspace*{-.6cm}
\caption{Limiting fragmentation in rapidity distributions at RHIC.}
\vspace*{-1.cm}
\label{feynman} 
\end{center}
\end{figure}

The most striking evidence in this sense comes
from the feature of the data known as  ``limiting fragmentation'' :
When plotted as functions of $y-y_{proj}$, the rapidity distributions 
measured at RHIC at {different} energies are to a good
approximation independent of energy (see Fig. 5; the final version of these
data, including the run at 19.6 GeV, is by now  available in Ref.
 \cite{PHOBOS}).
The meaning of this becomes clearer if one notices
that $y_{proj}-y\approx \ln(1/x)$, with 
$x$ the longitudinal momentum fraction of the produced particle.
Then, Fig. 5 shows that, with increasing energy, the ``fast'' (large $x$)
degrees of freedom do not change much, while new degrees of freedom
appear at smaller values of $x$. In other terms, the rapidity
distributions of the produced particles
are functions of $x$ alone, and not of the total energy:
This is similar to the Bjorken scaling of the parton distributions,
and is naturally explained by assuming that hadron 
interactions are short-ranged
in rapidity, so that the hadron distributions in rapidity
reproduce the corresponding distributions of the liberated partons.

\section{GLUON SATURATION}
\setcounter{equation}{0}

In order to solve the RGE (\ref{RGE}), one needs an initial condition
at low energy (say, at $x_0\simeq 10^{-1} \cdots 10^{-2}$). 
For a large nucleus, at least, it is reasonable to assume that the
only colour charges at such a large $x$ are the valence quarks,
which moreover are not correlated with each other, because they
typically belong to different nucleons \cite{MV94}. Then,
the initial weight function $W_{\tau_0}[\rho]$ is simply a gaussian,
with 2-point function \cite{MV94,Cargese}
\be\label{MV-corr}
\langle \rho_a(x_\perp)\rho_b(y_\perp)\rangle_0\,=\,
\delta_{ab}\delta^{(2)}(x_\perp-y_\perp)\,\mu_0,\qquad
\mu_0\equiv \,2\alpha_s A/R^2_A\,,\ee
which measures the colour charge
density of the valence quarks in the transverse plane.

When increasing $\tau$, the quantum evolution will
modify this simple initial condition by introducing correlations and 
non-linearities (that is, in general the evolved weight function will not be
a gaussian any longer \cite{SAT,path}). The details of the evolution
depend upon the structure of the kernel $\chi$ in eq.~(\ref{RGE}).
For the present purposes, it suffices to say that
$\chi$ depends upon
$\rho$ via the Wilson line (\ref{v}). This allows for an
intuitive distinction between the high-density and low-density regimes
\cite{SAT} :

--- At low energy, or large transverse momenta $k_\perp^2 \gg Q_s^2(\tau)$,
we are in a dilute regime where fields and sources are weak, and 
the Wilson lines can be expanded to lowest order:
$V^\dagger(x_{\perp})\approx 1 + ig \int dx^- A^+ (x^-,x_{\perp}) $.
Then, eq.~(\ref{RGE}) reduces to the BFKL equation for the
2-point function
$\langle \rho(k_\perp)\rho(-k_\perp)\rangle_\tau\equiv
\mu_\tau(k_\perp)$ \cite{JKLW97}, with solution (cf. eq.~(\ref{GBFKL})) :
\be\label{muBFKL}
 \mu_\tau(k_\perp)\,\simeq\, \sqrt{{\mu_0}{k_\perp^2}}\,
{\rm e}^{\omega \bar\alpha_s\tau}\,.\ee
As compared to eq.~(\ref{MV-corr}), we note the emergence of transverse
correlations, as well the rapid, {\it exponential}, increase with $\tau$.

--- At high energies, or low momenta $k_\perp^2 \simle Q_s^2(\tau)$,
the colour fields are strong, $A^+ \sim 1/g$, so the
Wilson lines rapidly oscillate and
average away to zero: $V\approx V^\dagger \approx 0$. Then the
kernel $\chi$ becomes independent of $\rho$, and the
r.h.s. of eq.~(\ref{RGE}) simplifies drastically \cite{SAT} : 
\be\label{RGElow}
{\del W_\tau[\rho] \over {\del \tau}}\,\approx\,
 {1\over 2\pi} \int_{k_\perp}\,\,k_\perp^2 \,
{\delta^2 \over {\delta
\rho_a(k_\perp)\delta
\rho_a(-k_\perp)}}\,\, W_\tau[\rho]\,.\ee
This is the standard diffusion equation for the Brownian motion. The
corresponding 2-point function increases only {\it linearly}
with the evolution ``time'' $\tau$ \cite{SAT,Cargese} :
\be\label{mu-sat}
\mu_\tau(k_\perp)
= \Big(\tau-\bar\tau(k_\perp)\Big)\frac{k_\perp^2}{\pi}\,
\,\simeq\,{k_\perp^2 \over \pi c \bar\alpha_s}\,
\ln{Q_s^2(\tau)\over k_\perp^2}\,,\ee
that is, {\it logarithmically} with the energy, since $\tau\sim
\ln s$. (In writing eq.~(\ref{mu-sat}), I
 have also used eq.~(\ref{Qs0}) for the saturation scale).
 For most purposes, a logarithm is as good
as a constant. We conclude that, at 
low momenta $k_\perp\ll Q_s(\tau)$, the colour
sources {\it saturate}, because of the non-linear effects
in the quantum evolution. 
\begin{figure} 
\begin{center} 
\centerline{\epsfig{file=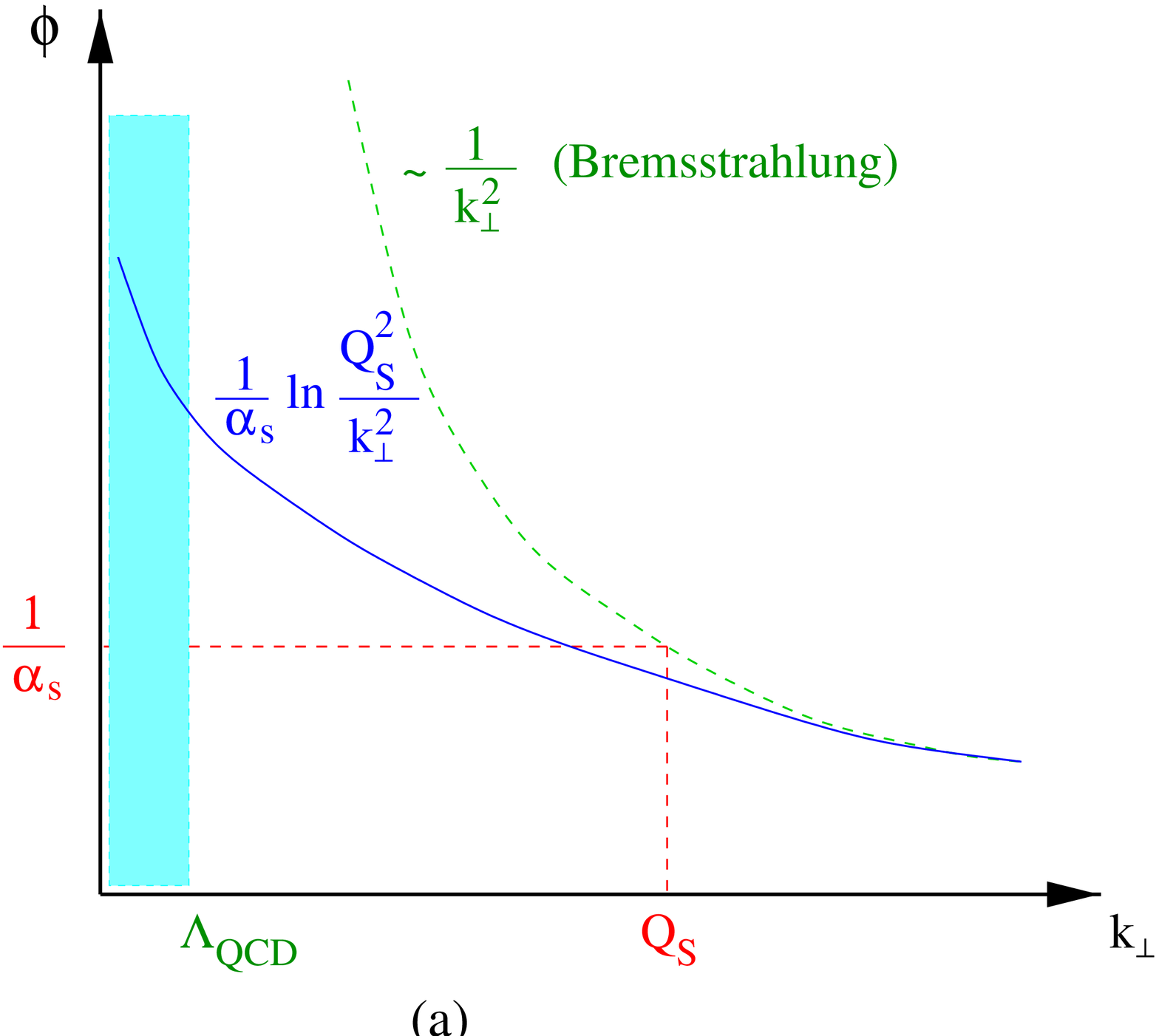,height=6.cm}
\epsfig{file=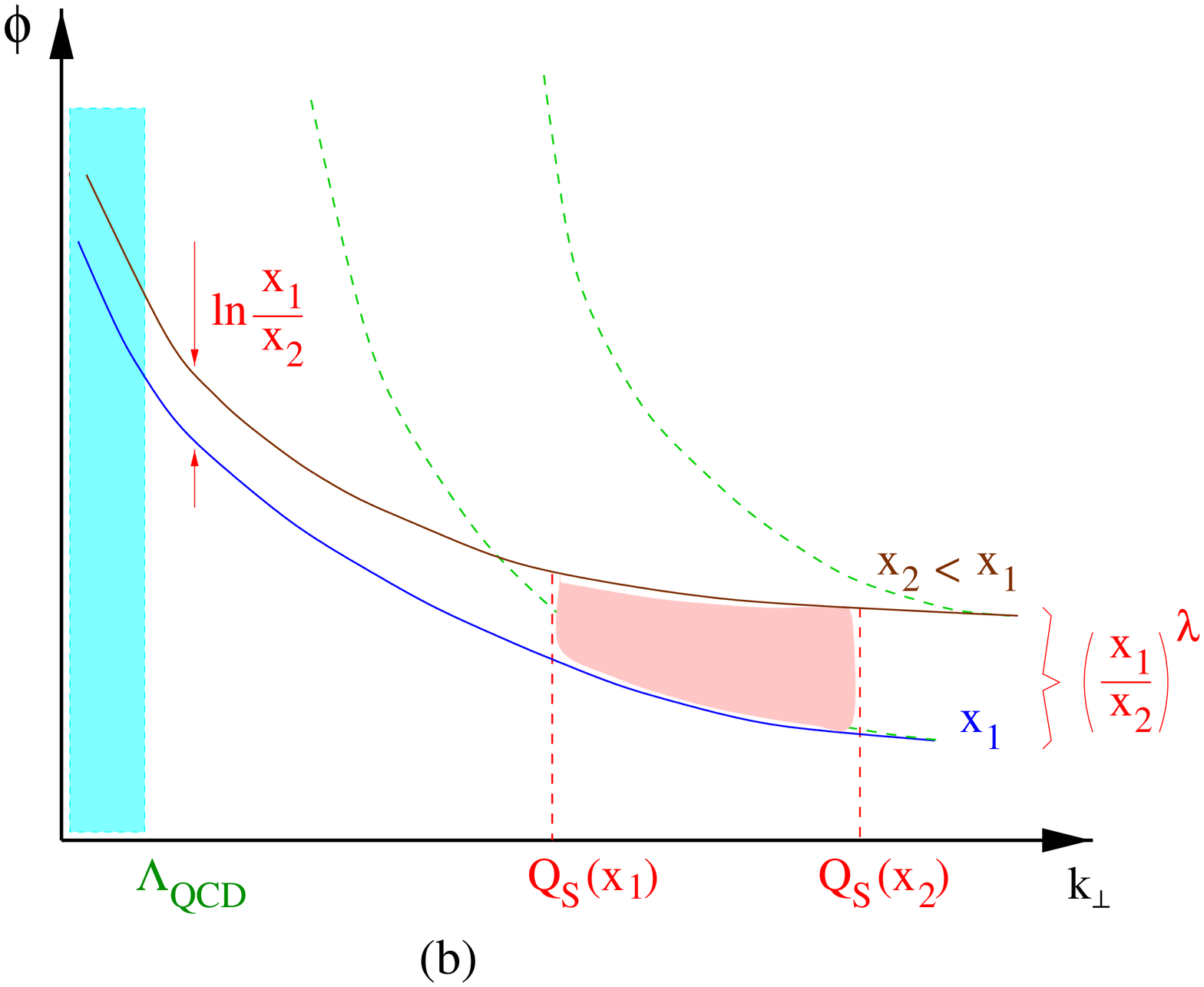,height=6.cm}}
\caption{(a) Gluon phase-space density as a function of $k_\perp$;
note the change in behaviour below $Q_s$.
 (b) The same as (a), but for two values of $x$.}
\end{center}\vspace*{-1.4cm}
\end{figure}

A second important feature of eq.~(\ref{mu-sat}) is that
it vanishes like $k_\perp^2$ when $k_\perp^2\to 0$. This means that
the saturated gluons screen each other in such a way
that {\it colour neutrality}
is achieved over a typical transverse area $1/Q_s^2(\tau)$.
This has important consequences for the infrared behaviour of the
perturbation theory, and also for the unitarization of cross-sections
at high energy \cite{FB} (see Sect. 7 below).

Gluon saturation can be seen also in the gluon distribution at low
$k_\perp$ \cite{AM2,SAT}. Let
\be\label{phi}
\phi_\tau(k_\perp)\,=\,\frac{1}{\pi R^2}\,
\frac{d^3 N}{d\tau d^2k_\perp}\,=\,\frac{1}{\pi R^2}\,
\frac{d \,x G(x,k_\perp^2)}{d^2k_\perp}\,
\ee
denote the gluon density in the transverse phase-space.
We have $\phi_\tau(k_\perp)\propto \mu_\tau(k_\perp)/k_\perp^2$
\cite{Cargese}.
This relation, together with the previous results for 
the charge-charge correlator 
$ \mu_\tau(k_\perp)$, implies the graphical representations
in Fig. 6, which are interpreted as follows:

At very large  $k_\perp \gg Q_s(\tau)$, one finds the expected
bremsstrahlung spectrum $\sim 1/k_\perp^2$, due to radiation
from independent colour sources (cf. eq.~(\ref{MV-corr})).
For lower momenta, but still well above $Q_s(\tau)$, the
spectrum gets softer $\sim 1/\sqrt{k_\perp^2}$ thanks to the BFKL evolution
(cf. eq.~(\ref{muBFKL})), but the density
increases rapidly with the energy,
as $x^{-\lambda}$ with $\lambda=\omega\bar\alpha_s$. Finally,
at $k_\perp \simle Q_s(\tau)$, there is {\it marginal saturation}
(in the sense of only a logarithmic increase) with respect
to both $1/k_\perp^2$ and $1/x$ (cf. eq.~(\ref{mu-sat})).
Thus, with decreasing $x$, the new gluons are predominantly produced 
at large transverse momenta $\simge  Q_s(\tau)$, where the density
is lower, and the repulsive interactions are less important.

\section{GEOMETRIC SCALING}
\setcounter{equation}{0}

At saturation, the gluon phase-space density defined in eq.~(\ref{phi})
reads (see Fig. 6) :
\be\label{SAT}
{\phi}_\tau(k_\perp)\,\simeq\,
{N^2_c-1\over 16 \pi^4}\,{1\over c \bar\alpha_s}\,
\ln{Q_s^2(\tau)\over k_\perp^2}\,.\ee
In addition to its saturation features alluded to before,
and to the $1/\alpha_s$ enhancement, this function exhibits
another interesting property, generally
referred to as ``geometric scaling'': It 
depends upon the two kinematical variables $\tau$ and
$k_\perp^2$ only via the ratio $k_\perp^2/Q_s^2(\tau)$ (the ``scaling
variable''). This is a natural property at saturation, where 
there is only one intrinsic scale, the saturation momentum.
Other quantities, like the dipole-hadron scattering amplitude
${\cal N}_\tau(r_{\perp})$, show in this regime 
a similar behaviour \cite{LT99,SAT}.

\begin{figure} 
\begin{center} 
\includegraphics[width=0.55\textwidth]
{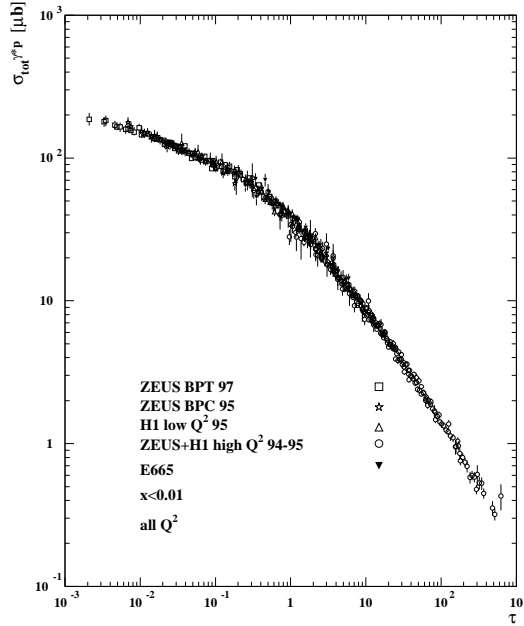} \vspace*{-.7cm}
\caption{HERA data on the cross section for $\gamma^*p$ DIS
from the region $x < 0.01$ and $Q^2 < 400{\rm GeV}^2$ plotted versus the 
scaling variable ${\cal T} = Q^2/Q^2_0(x)$. From Ref. \cite{geometric}. }
\label{gb} \vspace*{-.9cm}
\end{center}
\end{figure}
But at higher momenta $k_\perp^2\gg Q_s^2(\tau)$, or shorter distances
$r_{\perp} \ll 1/Q_s(\tau)$, where saturation effects should not
be important, there is a priori no reason to expect such a scaling.
And indeed, the gluon distribution in eqs.~(\ref{GBFKL}) or
(\ref{muBFKL}), or the scattering amplitude in the BFKL approximation
(which follows from eqs.~(\ref{nss}) and (\ref{GBFKL})) :
\be\label{NBFKL}
{\cal N}_\tau(Q^2)\Big|_{BFKL}\,\simeq\,
\exp\bigg\{ \omega\bar\alpha_s\tau - \frac{1}{2}\ln \frac{Q^2}{\Lambda^2}
-\frac{1}{2\beta\bar\alpha_s\tau}\bigg(\ln \frac{Q^2}{\Lambda^2}\bigg)^2
\bigg\}\ee
(with the notation $Q^2\equiv 1/r_{\perp}^2$)
show no obvious scaling. 

It may thus appear as a surprise that the HERA data on the total
cross section for $\gamma^*p$ deep inelastic scattering show nevertheless
geometric scaling to a quite good accuracy for $x < 0.01$  and all $Q^2$ up to 
relatively high values $\sim 400{\rm GeV}^2$ \cite{geometric,GS2} (see Fig.
\ref{gb}). Such $Q^2$ are significantly higher than the estimated value of
the saturation scale at HERA, as extracted from 
fits to $F_2$ within the ``saturation model'' \cite{GBW99} : 
$Q_s^2 \simeq 1 \cdots 2\, {\rm GeV}^2$.

Motivated by this experimental observation, we have reconsidered
the scaling properties of the dipole-hadron scattering amplitude
at momenta $Q^2$ above the saturation scale \cite{SCALING}, and discovered 
that the BFKL solution shows approximate scaling within
a window $Q_s^2 \le Q^2 \simle Q_s^4/\Lambda^2$. The upper
limit is of $O(100{\rm GeV}^2)$, in agreement with the data.

The basic argument is simple enough to be explained here:
Let me replace $\Lambda^2$ by $Q_s^2(\tau)$ as the reference scale
in eq.~(\ref{NBFKL}), with $Q_s^2(\tau)$ given by eq.~(\ref{Qs0}):
\be\label{LQS}
\ln (Q^2/\Lambda^2)\, = \,\ln (Q^2/Q^2_s(\tau))
+ c \bar\alpha_s\tau\,.\ee
This immediately yields, with $\lambda_s=1/2+c/\beta\approx 0.64$,
\be\label{BFKL_QS}
{\cal N}_\tau(Q^2)\,\simeq\,
\exp\Bigg\{- \lambda_s\ln \frac{Q^2}{Q_s^2(\tau)}
-\frac{1}{2\beta\bar\alpha_s\tau}
\left(\ln \frac{Q^2}{Q_s^2(\tau)}\right)^2
\Bigg\}.
\ee
Note that the exponent
in (\ref{BFKL_QS}) vanishes for ${Q^2}={Q_s^2(\tau)}$, as expected
(cf. eq.~(\ref{Nsat})).
From the equation above, we see that scaling emerges provided
$Q^2$ is close enough to $Q_s^2(\tau)$ (although still above it)
for the second term in the exponent to be negligible.
This provides the upper limit on $Q^2$ alluded to above.

To conclude this section, let me mention two recent applications
of the idea of geometric scaling at saturation to the phenomenology
of heavy ion collisions at RHIC. In Ref. \cite{SB01}, it has been argued
that the RHIC data for the
transverse momentum distributions of the produced hadrons show
geometric scaling in their dependences upon $m_T$ and centrality:
they are universal functions of $m_T/\Lambda_s(b)$, with
$b=$ the impact parameter. In Refs. \cite{KNL,KL}, it has been argued that
the scaling {\it violation} via the running of the coupling constant
[$\alpha_s\to\alpha_s(Q^2_s(\tau))$ in eq.~(\ref{SAT})]
may explain the observed centrality dependence of the multiplicity at RHIC.

\section{FROISSART BOUND FOR DIPOLE-HADRON SCATTERING}
\setcounter{equation}{0}

Let me now return to the dipole-hadron collision, and address the
fundamental question of the asymptotic behaviour of the total
cross-section as $s\to\infty$. Since this is 
obtained by integrating the scattering amplitude 
${\cal N}_\tau(x_{\perp},y_{\perp})\equiv {\cal N}_\tau(r_{\perp},b_{\perp})$
(with $r_\perp=x_\perp-y_\perp$  and $b_\perp=(x_\perp+y_\perp)/2$)
over all the impact parameters $b_\perp$ (see Fig. 8) :
\be\label{sigmadipole}
\sigma_{\rm tot}(\tau,r_\perp)\,=\,2\int d^2b_\perp\,
{\cal N}_\tau(r_{\perp},b_{\perp})\,,\ee
it is clear that there are actually two issues that are
concerned by this question: \\ ({\it i}) the increase of
the scattering amplitude with $\tau$ 
at fixed impact parameter, and ({\it ii})
the decrease of the scattering amplitude with $b_{\perp}$
at large impact parameters.
The first issue is that of {\it unitarity} : the local  
scattering amplitude cannot exceed the unitarity bound 
${\cal N}_\tau(r_\perp,b_{\perp})\le 1$. The second issue is
that of {\it confinement}: this is why hadron interactions
have only a finite range.
\begin{figure} 
\begin{center} 
\includegraphics[width=0.7\textwidth]
{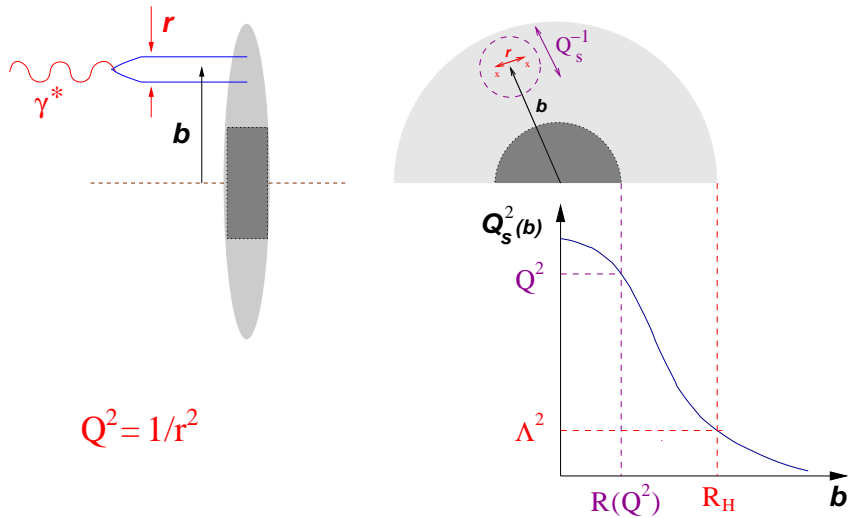} \vspace*{-.6cm}
\caption{Dipole-hadron scattering in longitudinal (left) and transverse
(right) projections.}
\end{center}\vspace*{-.8cm}
\label{GREY}
\end{figure}

({\it i}) The {\it unitarization} problem can be addressed within the
perturbative framework. Clearly, the BFKL equation violates unitarity,
but this is restored by saturation, which ensures that
${\cal N}_\tau(r_\perp,b_{\perp})\simeq 1$ for any $r_\perp\simge
1/Q_s(\tau,\bb)$. Here, $Q_s(\tau,b_{\perp})$ is the {\it local}
saturation scale, defined as in eq.~(\ref{Qs}), but with
${x G(x,Q^2)}/{\pi R^2}$ replaced by $x G(x,Q^2,\bb) 
\equiv {dN}/{d\tau d^2\bb}$ (the gluon density at the given $\bb$).

The behaviour of ${\cal N}_\tau(r_\perp,b_{\perp})$ as
a function of $r_\perp$ at fixed $\bb$ is illustrated in Fig. 9 for
two different energies. One sees a change in regime around
$\rr\sim 1/Q_s(\tau,\bb)$ from ``colour transparency'' at small $r_\perp$
(cf.  eq.~(\ref{nss})) to ``blackness'' at large $r_\perp$.
This behaviour can be obtained by numerically 
solving the BK equation (\ref{BK}) \cite{LT99,AB01,Motyka}.
\begin{figure} 
\begin{center} 
\includegraphics[width=0.35\textwidth]
{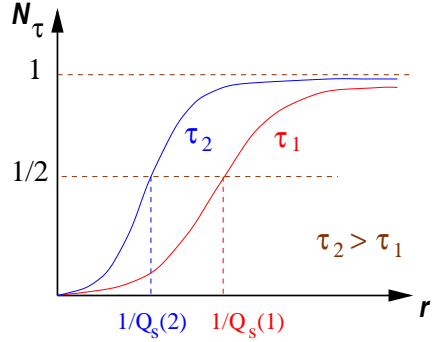} \vspace*{-.5cm}
\caption{The scattering amplitude as a function of $\rr$ 
for two values of $\tau$.}
\end{center}\vspace*{-1.cm}
\label{NR}
\end{figure}
Some limiting regimes can be also studied analytically.
For $r_\perp\ll 1/Q_s(\tau,\bb)$, one obtains \cite{AM0,SAT}.:
\be\label{GM-DIP}
{\cal N}_\tau(r_\perp,\bb)\,\simeq\,1-{\rm exp}\bigg\{-\rr^2
\frac{\pi^2\alpha_s C_F}{N_c^2-1}
\,\,x G(x,1/\rr^2,\bb)\bigg\},\ee
which exhibits the unitarizing role of the
multiple scattering (compare to eq.~(\ref{nss}), which describes a single
scattering). For $r_\perp\gg 1/Q_s(\tau,\bb)$ one rather has \cite{LT99,SAT} :
\beq\label{Nscaling}
{\cal N}_\tau(\rr,b_{\perp})\,\approx\, 1\,-\,\exp\bigg\{-
\frac{1}{2c}\Big(\ln {Q^2_s(\tau,b_{\perp})}{\rr^2}\Big)^2\bigg\}\,,
\eeq
which shows geometric scaling with the scale set by the local
saturation scale $Q^2_s(\tau,b_{\perp})$.  

With increasing $\tau$ at fixed $\bb$, 
the saturation scale $Q^2_s(\tau,b_{\perp})$
increases and eventually becomes larger than the external
resolution scale $Q^2\equiv 1/\rr^2$ (see Fig. 8). 
Once this happens, the local scattering amplitude reaches the unitarity limit:
${\cal N}_\tau(r_\perp,b_{\perp})\simeq 1$ 
(i.e., the hadron becomes locally ``black''), and then remains
constant when the energy is further increased. 
Since the gluon density is larger at the center of the hadron,
the black disk first appears at $\bb=0$, and then extends radially
with increasing $\tau$ (see below). This transverse expansion explains why the
total cross-section keeps growing even at very high energies.
Let $R(\tau, Q^2)$ denote the {\it black disk radius}. Then, by definition :
\be\label{INT_sat}
{\cal N}_\tau(Q^2,\, \bb)\,\simeq\, 1\qquad 
{\rm for}\qquad b\,\le\,R(\tau,Q^2)
\ee
[with ${\cal N}_\tau(Q^2,\, \bb)\equiv  {\cal N}_\tau(r_\perp=1/Q,b_{\perp})$
from now on], or, equivalently,
\be\label{condR}
Q_s^2(\tau,\bb)\,=\,Q^2\qquad{\rm for}\qquad b\,=\,R(\tau, Q^2).\ee
In order to compute $R(\tau,Q^2)$,
 one needs to understand the evolution of the scattering amplitude
with $\tau$ at impact parameters {\it outside} the black disk, i.e.,
in the ``grey area'' at $b>R(\tau, Q^2)$ (see Fig. 8). This brings us
to the problem of {\it confinement}.

({\it ii}) If $b > R(\tau, Q^2)$, 
the gluon density is low and ${\cal N}_\tau(Q^2,\, \bb)$ is simply proportional
to the local distribution function  $x G(x,Q^2,\bb)$, cf.  eq.~(\ref{nss}).
That is, the dipole is probing the gluon fields with momenta $\kk^2\le Q^2$
created at $\bb$ by all the colour sources in the hadron. Because of confinement,
such fields must fall off exponentially when the transverse separation
between $\bb$ and the colour sources is larger than the pion Compton wavelength
$1/m_{\pi}$ (since $m_{\pi}$ is the lowest mass gap in QCD). We thus expect 
\cite{FB} :
\be\label{Nexp}
{\cal N}_\tau(Q^2,b_{\perp})\,\propto \,{\rm e}^{-2m_{\pi}b}\,
\qquad {\rm for}\qquad b- R(\tau, Q^2) \simge 1/2m_{\pi}\,,\ee
where {\it twice} the pion mass 
enters the exponent because of isospin conservation:
The long-range scattering is controlled by pion exchange, and pions
have isospin one, while gluons have isospin zero, so the scattering
proceeds via the exchange of (at least) two pions.
A similar exponential decrease with $\bb$ holds for the
saturation scale \cite{FB} (see Fig. 8).

Since ${\cal N}_\tau(Q^2,\, \bb)$ is rapidly decreasing at 
$\bb\gg R(\tau, Q^2)$, the total cross-section is dominated by the 
black disk:
\be\label{sigmaBD}
\sigma_{\rm tot}(\tau,Q^2)\,\simeq \,
2\pi R^2(\tau, Q^2).\ee
Thus, the relevant question is: 
How fast expands the black disk with $\tau$ ? 

As shown in Ref. \cite{FB}, and I shall briefly explain here, one can 
answer this question by combining the perturbative (but {\it non-linear})
 evolution with $\tau$ with the non-perturbative
boundary condition (\ref{Nexp}) due to confinement.
Perturbation theory applies since the points at which the hadron turns from
``grey'' to ``black'' (see Fig. 8) are characterized by sufficiently large
gluon densities, as we shall see.

Specifically, we are in a weak coupling regime provided the local
saturation scale is hard: $Q_s^2(\tau,b) \gg \Lambda^2_{QCD}$.
Given eq.~(\ref{condR}) and the fact that the dipole is small
($Q^2\gg \Lambda_{QCD}^2$), it is clear that there exists a
{\it grey corona} at $R(\tau,Q^2)< b < R_H(\tau)$ within which
perturbation theory applies (since $\Lambda^2_{QCD}\ll Q_s^2(\tau,b)\ll Q^2$ 
at any $\bb$ within this corona). Here, $R_H(\tau)$ is the radial distance
at which the saturation scale falls down
to $\Lambda_{QCD}$ (see Fig. 8).
We shall check a posteriori that this perturbative corona is wide enough
to allow for the calculation of the {\it expansion rate} of the black disk.

For impact parameters within this corona, 
${\cal N}_\tau(Q^2,b_{\perp})\ll 1$, so one may expect the
linear, BFKL, approximation to apply. This turns out to be correct
eventually, but the complete argument is more subtle \cite{FB}: 
By itself, the BFKL
equation would allow for long-range interactions between the incoming
dipole and the colour sources in the hadron, but in the full equation
(\ref{BK}) such interactions are suppressed by the non-linear effects 
whenever the transverse separation exceeds $ 1/Q_s(\tau,\bb)$.
Physically, this reflects the fact that the saturated gluons 
are {\it colour neutral} (cf. eq.~(\ref{mu-sat})),
and therefore couple to the incoming $q\bar q$ pair only via 
dipole-dipole forces, which have a rapid fall-off with $\bb$,
and therefore do not
contribute significantly to the scattering in the grey corona. Rather,
the dominant contributions come from the {\it nearby} colour sources, i.e., 
the sources located
within a saturation disk around the impact parameter $ \bb$ (see Fig. 8):
\be\label{SR}
|z_\perp-\bb|\,\ll\, {Q_s^{-1}(\tau,\bb)}
\,.\ee
To describe such short-range
contributions, one can indeed rely on the BFKL equation, but which is
supplemented with  an infrared cutoff $\sim Q_s(\tau,\bb)$,
to remove the unphysical long-range contributions.

The crucial point is that the long-range interactions are cut off by
saturation effects already at the {\it short} scale $ 1/Q_s(\tau,\bb)$,
which is much shorter than the soft scale $1/ \Lambda_{QCD}$ 
at which the confinement plays a role. 
Because of that, the perturbative evolution with $\tau$ is not 
sensitive to the transverse inhomogeneity in the hadron, which manifests
itself only on the very large scale $1/m_{\pi}\sim 1/ \Lambda_{QCD}$ 
(cf. eq.~(\ref{Nexp})). That is, the evolution proceeds 
{\it quasi-locally} in the impact parameter
space, in such a way that the $\bb$--dependence of
the scattering amplitude {factorizes out}, and is fixed by
the initial conditions. To summarize:
\be\label{BFKL_sol1}
{\cal N}_\tau(Q^2,\bb)\Big|_{grey}
&\simeq&{\cal N}_\tau(Q^2)\Big|_{BFKL}
\times\, {\rm e}^{-2m_{\pi}b}\nonumber\\
&\simeq&\exp\left\{
-2m_\pi b + \omega\bar\alpha_s\tau - \frac12\ln \frac{Q^2}{\Lambda^2}
\right\},
\ee
where the $\bb$--dependence comes from eq.~(\ref{Nexp}), and 
in writing the second line I have used the BFKL solution (\ref{NBFKL}) 
(with the diffusion term omitted, as appropriate at high energies).
This applies only for points in the grey area ($b>R(\tau,Q^2)$), but
it can be used to estimate the black disk radius from the
condition (\ref{INT_sat}). One obtains:
\be\label{BD_asy}
R(\tau,Q^2)\, \simeq \, \frac{1}{2m_\pi}\left( \omega \bar\alpha_s \tau 
-\frac12 \ln \frac{Q^2}{\Lambda^2}\right),
\ee
which implies the following result for the
total cross-section at high energy \cite{FB}  :
\be\label{sigmaFC}
\sigma_{\rm tot}(s,Q^2)\,\approx\,
\frac{\pi}{2}\left(\frac{\omega \bar\alpha_s}{m_{\pi}}\right)^2\ln^2s\qquad 
{\rm as\,\,\,\,} s
\to\infty.\ee
This {\it saturates} the Froissart bound, with a proportionality
coefficient which is {\it universal} (i.e., the same for all hadrons), and 
which reflects the combined role of perturbative and
non-perturbative physics in controlling the asymptotic behaviour
 at high energy.

At this point, one should recall that the Froissart bound 
$\sigma_{\rm tot}\le \sigma_0 \ln^2s$ \cite{Froissart} 
is a consequence of general principles
(unitarity, crossing, and analiticity), but does not rely on  
detailed dynamical information. So, in reality, this bound may very well 
be not saturated. It so happens, however, that the measured cross-sections 
(e.g., for $pp$ and $p \bar p$ scattering) show a slow, but monotonous,
increase with $s$, which can be reasonably well fitted by a $\ln^2 s$ 
behaviour \cite{BNetal}. This suggests that the Froissart bound is actually
saturated in nature, and the mechanism described above
provides a physical picture for such a saturation, with a definite prediction
for the scale $\sigma_0$.

From eq.~(\ref{BFKL_sol1}), one can compute also the radius
$R_H(\tau)$ where the saturation scale decreases
to $\Lambda_{QCD}$ (beyond which perturbation theory fails to apply).
According to eqs.~(\ref{INT_sat})--(\ref{condR}), this is the same 
as the black disk radius for a large dipole with $Q^2\sim\Lambda_{QCD}^2$:
\be\label{RH1}
R_H(\tau)\,\approx\,\frac{\omega \bar\alpha_s }{2m_\pi}\, \tau \, .\ee
One can now compute the radial extent of the grey corona within
which perturbation theory is applicable:
\beq\label{RR}
R_H(\tau)\,-\,R(\tau,Q^2)\,\approx\,\frac{1}{4m_\pi}\, 
\ln \frac{Q^2}{\Lambda^2}\,.
\ee
This is independent of $\tau$, and much larger than $1/m_\pi$ (because
of the large logarithm $\ln(Q^2/\Lambda^2)$), which
 demonstrates the consistency of the previous calculation:
During a ``time'' interval $\bar\alpha_s \Delta \tau \sim 1$ 
(the typical rapidity increment at high energy), the 
black disk expands from $R(\tau,Q^2)$ to $R(\tau,Q^2)+\omega/2m_\pi$,
cf. eq.~(\ref{BD_asy}), which is much smaller than $R_H(\tau)$,
and therefore still in the region controlled by perturbation theory.
That is, the expansion of the black disk proceeds within
the perturbative corona for intervals $\Delta\tau$ which are large enough
to allow for a controlled calculation of the  rate of this expansion
\cite{FB}.

On the other hand, one cannot rely on perturbation theory alone
to follow the expansion of the black disk for {\it arbitrarily} large ``time''
intervals $\Delta\tau$. For instance, it makes physically no sense to
compute in perturbation theory the gluon distribution or the
scattering amplitude at very large impact parameters $\bb\gg R_H(\tau)$,
where formally $Q_s^2(\tau,b)\ll\Lambda^2_{QCD}$. But if one insists
in doing so, one finds that, because of the lack of confinement, the
perturbative evolution (say, according to the BK equation (\ref{BK}))
generates long-range gluon fields which replace,
at sufficiently large distances, the exponential fall-off of the
initial distribution by just a power-law fall-off. 
If one pushes the perturbative expansion until the black disk
enters this power-law tail, then its expansion rate speeds up,
and eventually violates unitarity \cite{KW02}. 

Clearly, this violation is 
an artifact of pushing perturbation theory beyond its limits of validity. 
To generate a significant power-law tail, the perturbative fields
must propagate over distances many times the pion Compton wavelength.
Such long-range  fields are certainly unphysical: 
in the real world, they are removed by confinement.

The true difficulty of perturbation theory is that
it cannot be used to {\it generate} the exponential tail
at large distances. But this is to be expected: 
such a tail can arise only from confinement.
In any case, this difficulty has no incidence on 
the previous calculation of the {\it rate} of the expansion,
since this requires only a limited evolution in $\tau$ which is driven
by short-range interactions. The prediction (\ref{sigmaFC}) of this
 calculation can be extended\footnote{Indeed, at any
$\tau$, one can repeat this calculation and obtain an expansion
rate consistent with eq.~(\ref{BD_asy}).}
to arbitrarily large $\tau$,
although the perturbative evolution becomes meaningless eventually.

To conclude, while perturbation theory alone appears to be sufficient to
describe unitarization at fixed impact parameter, one still needs some
information about the finite range of the strong interactions
in order to be able to compute total cross-sections. This is reminiscent of 
an old argument by Heisenberg \cite{Heisenberg} which combines unitarity and 
short-rangeness (as modelled by a Yukawa potential) to deduce
cross-sections which saturate the Froissart bound. Fifty years later,
our progress in understanding high energy strong interactions allows us
to confirm Heisenberg's intuition, and identify short-rangeness with
confinement, and unitarization with saturation.

\section{APPLICATIONS TO PHENOMENOLOGY}

I conclude this review with a succint enumeration of recent applications
of the concept of CGC to the phenomenology of deep inelastic scattering 
at HERA, and of relativistic heavy ion collisions at RHIC (with 
perspectives for LHC). More detailed discussions and more
references can be
found in \cite{QM02}. But let me start with a few words of caution:

({\it a\,}) The theoretical analysis underlying this concept relies heavily 
on the smallness of the coupling constant, which in turn requires a rather 
hard saturation scale: $Q_s^2(\tau) \gg \Lambda^2_{QCD}$. But
for the currently available energies this condition is only
marginally satisfied. Indeed, one estimates
 that $Q_s^2\sim 1\cdots 2 \,{\rm GeV}^2$
at both RHIC and HERA \cite{QM02,GBW99}.

({\it b\,}) The general formalism, and most of the theoretical calculations,
have been developed so far only to lowest order in $\alpha_s$. But for realistic
comparisons with phenomenology, a NLO formalism (at least) is necessary.
The NLO corrections to the BFKL kernel became finally available
\cite{NLBFKL}, but it turned out that resummation techniques are necessary 
to render these corrections meaningful \cite{Salam99}. 
Very recently, these methods have been used to 
estimate the saturation scale from the NLO BFKL equation \cite{T02}
(via the same strategy as already used at LO; cf. Sect. 2.2).
As a result, the logarithmic derivative $\lambda_s \equiv d \ln[Q_s^2(\tau)
/\Lambda^2]/d\tau$ appears to decrease from the LO estimate
$c\bar\alpha_s \simeq 1$ (cf. eq.~(\ref{Qs})) to 
$\lambda_s \simeq 0.30$, which is a reasonable value
for the HERA phenomenology \cite{GBW99}.
A fully non-linear NLO formalism is still to be developed
(see however \cite{BB01}).

({\it c\,}) For applications to heavy ion collisions, the present
formalism provides only the {\it initial conditions}, i.e., the 
wavefunctions of the incoming nuclei. Based on this, 
various methods have been developed to compute the melting of the CGC 
in the early stages of the collision, and the associated gluon production
\cite{KM98,KV,YK01,KNL}. But very little
is known from first principles about the subsequent evolution of the
liberated partons. Their initial spectra and distributions
can, and most certainly will,
be modified by final state interactions, flow, or thermalization, 
and may further change in the process of hadronization,
but these various processes are not yet under theoretical control.
It is therefore difficult, if not impossible,
to unambiguously identify at this stage the effects of the initial
conditions (say, of saturation) in the measured particle yields 
and correlations.

This being said, it is remarkable that there exists already 
a significant amount of data, both at HERA and at RHIC, which hint
towards saturation and the CGC, or at least are consistent with this
physical picture.

Let me start with DIS at small-$x$, for which the theoretical calculations
are better under control, and quantitative
predictions can be made. It has been shown in
Refs. \cite{GBW99} that the HERA data for both inclusive and
diffractive proton structure functions can be well accounted for by
a phenomenological model which incorporates saturation.  The success
of this simple model for {\it diffraction} is particularly significant, 
since the diffractive cross-section is more sensible than the
inclusive one to dipole fluctuations with a large transverse size $r_\perp$, 
and thus to saturation effects in the proton wavefunction. 
The same model has motivated the search for geometrical
scaling in the HERA data \cite{geometric}, 
with the remarkable results discussed in Sect. 6. 
As shown in Ref. \cite{Munier01}, it is possible to use the data on the
differential cross-section for vector meson production in DIS
(say, $\gamma^* p \to\rho p$) to obtain information about the 
impact parameter dependence of the dipole scattering amplitude.
The phenomenological analysis in Ref. \cite{Munier01} indicates a reasonable
amount of blackness in the proton at central impact parameters already
for $x\sim 10^{-3}$.

Consider heavy ion collisions now. In  Refs. \cite{KNL,KL}, the particle
production in $Au-Au$ collisions at RHIC 
has been first analyzed from the perspective of the CGC.
Specifically, the  measured hadron multiplicities have been compared, in so far
as their dependences upon centrality, rapidity, and total energy are concerned,
to the respective predictions of the CGC picture. The agreement is quite good,
in fact, remarkably good, given the many sources of uncertainty that
I have mentioned before.

A  systematic approach, aiming at the numerical resolution
of the classical Yang-Mills equations which describe the scattering of two CGCs, 
has been developed and successively refined in Refs. \cite{KV}. In this approach 
too, one finds a reasonable agreement between the centrality dependence of the
gluons liberated in the collision (as obtained from lattice simulations) 
and that of the charged hadrons measured at RHIC.

For proton-nucleus ($pA$) collisions, and for DIS, there
are also analytic calculations of the gluon and particle production,
which include all multiple rescatterings off the strong colour field
(the CGC) of the nucleus. The inclusive gluon production cross-section
has been computed for  $pA$ \cite{KM98,YK01,DM01}
and DIS \cite{KT01}, and a similar calculation for $AA$ has been attempted
in \cite{YK01}. Still for $pA$ collisions, one has calculated the 
cross-sections for the production of jets
\cite{DJ01}, photons and dileptons \cite{GJ01}.
The charm production from the CGC in peripheral heavy-ion collisions
has been investigated in \cite{GP01}. 
See also the closely related work in Ref. \cite{Boris99}.
Instantons in the saturation environment have been considered in
Ref. \cite{KKL01}. 
At a conceptual level, there has been significant
progress towards understanding the role of inelastic scattering
for thermalization, and the way how kinetic theory could emerge
from the classical field dynamics that applies 
at the earliest stages  \cite{bottom}.


\begin{thebibliography}{9}

\bibitem{BFKL}
L.N.~Lipatov, {\it Sov.\ J.\ Nucl.\ Phys.}\,{\bf 23} (1976) 338;
E.A.~Kuraev, L.N.~Lipatov and V.S.~Fadin, {\it Zh. Eksp.
Teor. Fiz}
{\bf 72}, 3 (1977) ({\it Sov. Phys. JETP }{\bf 45} (1977) 199);
Ya.Ya.~Balitsky and L.N.~Lipatov, {\it Sov.\ J.\ Nucl.\ Phys.} {\bf 28}
(1978) 822.

\bibitem{GLR}   L.V.~Gribov, E.M.~Levin, and M.G.~Ryskin, {\it Phys.
Rept. } {\bf 100} (1983) 1.
 
\bibitem{MQ86}  A.H.~Mueller and J.~Qiu, {\it Nucl. Phys.} {\bf
B268} (1986)  427.

\bibitem{BM87}
J.-P. Blaizot and A. H. Mueller, {\it Nucl. Phys.} {\bf B289} (1987) 847.

\bibitem{MV94}
L. McLerran and R.~Venugopalan, {\it Phys.\ Rev.}\ {\bf D49} (1994) 2233;
{\it ibid.} {\bf 49} (1994) 3352; {\it ibid.} {\bf 50} (1994) 2225.

\bibitem{K96}
Yu.V.~Kovchegov, {\it Phys.\ Rev.}\ {\bf D54} (1996), 5463;
 {\it Phys.\ Rev.}\ {\bf D55} (1997), 5445.


\bibitem{JKMW97}
J.~Jalilian-Marian, A.~Kovner, L.~McLerran, H.~Weigert,
{\it Phys.\ Rev.}\ {\bf D55} (1997) 5414.

\bibitem{KM98}
Yu.V.~Kovchegov and A.H.~Mueller, {\it Nucl.\ Phys.} {\bf B529}
 (1998), 451.

\bibitem{LM00} 
C. S. Lam and G. Mahlon, {\it Phys. Rev.} {\bf D62} (2000) 114023;
{\it ibid.} {\bf D64} (2001) 016004.

\bibitem{JKLW97}
J.~Jalilian-Marian, A.~Kovner, A.~Leonidov and  H.~Weigert,
{\it Nucl.\ Phys.}\ {\bf B504} (1997) 415;
{\it Phys.\ Rev.}\ {\bf D59} (1999) 014014.

\bibitem{PI}
E.~Iancu, A.~Leonidov and L.~McLerran,
 {\it Nucl. Phys.}~{\bf A692} (2001), 583;
{\it Phys. Lett.} {\bf B510} (2001) 133;
E.~Ferreiro et al., 
{\it Nucl. Phys.} {\bf A703} (2002) 489.


\bibitem{Cargese}
E.~Iancu, A.~Leonidov and L.~McLerran,
{\it The Colour Glass Condensate: An Introduction}, hep-ph/0202270.
Lectures given at the NATO Advanced Study
Institute ``QCD perspectives on hot and dense matter'', August 6--18, 2001,
Carg\`ese, France.

\bibitem{AM2} A. H. Mueller, {\it Nucl. Phys.} {\bf B558} (1999) 285.

\bibitem{SAT}
E.~Iancu and L.~McLerran, {\it Phys. Lett.} {\bf B510} (2001) 145.

\bibitem{geometric}
A.M.~Sta\'sto, K.~Golec-Biernat, and J.~Kwieci\'nski, 
{\it Phys. Rev. Lett.} {\bf 86} (2001) 596.

\bibitem{GS2}
D. Schildknecht, B. Surrow, and M. Tentyukov, {\it Phys. Lett.} {\bf B499} (2001)
 116.

\bibitem{SCALING}
E.~Iancu, K. Itakura, and L. McLerran, {\it Nucl. Phys.} {\bf A708} (2002) 327.

\bibitem{FB}
E.~Ferreiro, E.~Iancu, K. Itakura, and L.~McLerran, 
 {\it Nucl. Phys.} {\bf A710} (2002) 373.

\bibitem{QM02}
See the contributions by R. Baier, D. Kharzeev, A. Krasnitz, A. Mueller, and
H. Satz in this volume.

\bibitem{AM0}
A. H. Mueller, {\it Nucl. Phys.} {\bf B335} (1990) 115;
N.N. Nikolaev and B.G. Zakharov, {\it Z. Phys.} {\bf C49}
(1991) 607, {\it ibid.} {\bf C53} (1992) 331.

\bibitem{Froissart} M.~Froissart, {\it Phys. Rev. } {\bf 123} (1961) 1053.

\bibitem{BH}
W. Buchmuller, M.F. McDermott and A. Hebecker, 
{\it Nucl. Phys.} {\bf B487} (1997) 283;
 W. Buchmuller, T. Gehrmann, and A. Hebecker
{\it Nucl. Phys.} {\bf B537} (1999) 477.

\bibitem{B}
I.~Balitsky, {\it Nucl.\ Phys.}\ {\bf B463} (1996) 99;
hep-ph/0101042.


\bibitem{path}
J.-P. Blaizot, E.~Iancu, and H. Weigert, hep-ph/0206279, to appear
in {\it Nucl. Phys.} {\bf A}. 

\bibitem{RW}
K. Rummukainen and H. Weigert, in preparation.

\bibitem{K}  Yu. V. Kovchegov, {\it Phys. Rev.} 
{\bf D60} (1999), 034008; {\it ibid.} {\bf D61} (2000) 074018.

\bibitem{B00}
M. Braun, {\it Eur. Phys. J.} {\bf C16} (2000) 337.

\bibitem{W}
H. Weigert, {\it Nucl. Phys.} {\bf A703} (2002) 823.

\bibitem{PHOBOS}
B.B. Back, et al. (PHOBOS collaboration), nucl-ex/0210015.


\bibitem{LT99} E.~Levin and K.~Tuchin,
{\it  Nucl. Phys.} {\bf B573} (2000) 833;
{\it Nucl. Phys.} {\bf A691} (2001) 779.

\bibitem{GBW99}
K. Golec-Biernat, M. W\"usthoff, {\it Phys. Rev.} {\bf D59} (1999)
014017; {\bf D60} (1999) 114023.

\bibitem{SB01}
J. Schaffner-Bielich, D. Kharzeev, L. McLerran and R. Venugopalan,
{\it Nucl. Phys.} {\bf A705} (2002) 494.

\bibitem{KNL}
D. Kharzeev and M. Nardi, {\it Phys. Lett.} {\bf B507} (2001) 121.

\bibitem{KL}
D. E. Kharzeev and E. Levin, {\it Phys. Lett.} {\bf  B523} (2001) 79.


\bibitem{AB01}
N.~Armesto and M.~Braun, {\it Eur. Phys. J.} {\bf C20} (2001) 517;
{\it ibid.} {\bf C22} (2001) 351. 

\bibitem{Motyka} K.~Golec-Biernat, L.~Motyka, and A.M.~Sta\'sto, 
{\it Phys. Rev.} {\bf D65} (2002) 074037. 

\bibitem{BNetal}
J.R. Cudell et al, {\it Phys. Rev. } {\bf D65} (2002) 074024.

\bibitem{KW02}
A. Kovner and U.A. Wiedemann,   hep-ph/0112140; hep-ph/0204277; hep-ph/0207335.

\bibitem{Heisenberg}W.~Heisenberg, {\it Z. Phys.} {\bf 133} (1952) 65.

\bibitem{NLBFKL} V.S.~Fadin and L.N.~Lipatov, 
{\it Phys. Lett.} {\bf B429} (1998) 127;
G. Camici and M. Ciafaloni, {\it Phys. Lett.} {\bf B430}  (1998) 349.

\bibitem{Salam99}
G.P. Salam, {\bf JHEP 9807} (1998) 19; M. Ciafaloni, D. Colferai, 
 {\it Phys. Lett.} {\bf B452} (1999) 372; M. Ciafaloni, D. Colferai, 
and G.P. Salam, {\it Phys. Rev. } {\bf D60} (1999) 114036.

\bibitem{T02}
D.N. Triantafyllopoulos, hep-ph/0209121.

\bibitem{BB01}
I.~Balitsky and A.V. Belitsky, {\it Nucl. Phys.} {\bf B629} (2002) 290.


\bibitem{Munier01}
S. Munier,  A.M.~Sta\'sto, and A.H. Mueller, {\it Nucl. Phys.} {\bf B603}
(2001) 427.


\bibitem{KV}
A. Krasnitz, R. Venugopalan, {\it Phys. Rev. Lett.} {\bf 84}
(2000) 4309;  {\it ibid.}  {\bf 86} (2001) 1717;
A. Krasnitz, Y. Nara, and  R. Venugopalan, {\it ibid.}
{\bf 87} (2001) 192302; hep-ph/0209269.


\bibitem{YK01}
Yu. V. Kovchegov,  {\it Nucl.Phys.} {\bf A692} (2001) 557.

\bibitem{DM01}
A. Dumitru and L. McLerran, {\it Nucl. Phys.} {\bf A700} (2002) 492.

\bibitem{KT01}
Yu. V. Kovchegov, {\it Phys.Rev.} {\bf D64} (2001) 114016; Yu. V. Kovchegov
and K.~Tuchin, {\it ibid.}  {\bf D65} (2002) 074026.

\bibitem{DJ01}
A. Dumitru and J.~Jalilian-Marian, {\it Phys. Rev. Lett.} {\bf 89}
 (2002) 022301.

\bibitem{GJ01}
F. Gelis and J.~Jalilian-Marian, {\it Phys. Rev.} {\bf D66} (2002) 014021;
hep-ph/0208141.

\bibitem{GP01}
 F. Gelis and A. Peshier, {\it Nucl. Phys.} {\bf A697} (2002) 879;
{\it ibid.} {\bf A707} (2002) 175.

\bibitem{Boris99}
B.Z. Kopeliovich, A. Schaefer, and A.V. Tarasov,  
{\it Phys.Rev.} {\bf C59} (1999) 1609; 
B.Z. Kopeliovich, J. Nemchik, A.Schaefer and  A.V. Tarasov,
{\it Phys.Rev.} {\bf C65} (2002) 035201;
B.Z. Kopeliovich, A.V. Tarasov, and J. Huefner, {\it Nucl. Phys.} {\bf A696}
 (2001) 669; B.Z. Kopeliovich and A.V. Tarasov,
{\it Nucl. Phys.} {\bf A710} (2002) 180.

\bibitem{KKL01}
D. E. Kharzeev, Yu. V. Kovchegov, and E. Levin,
{\it Nucl. Phys.} {\bf A699} (2002) 745.

\bibitem{bottom}
R. Baier, A.H. Mueller, D. Schiff and D.T. Son, {\it Phys. Lett.} {\bf B502}
(2001) 51.


\end{thebibliography}
\end{document}